\documentstyle{mn}
\input epsf

\def\spose#1{\hbox to 0pt{#1\hss}}
\def\simlt{\mathrel{\spose{\lower 3pt\hbox{$\mathchar"218$}}
     \raise 2.0pt\hbox{$\mathchar"13C$}}}
\def\simgt{\mathrel{\spose{\lower 3pt\hbox{$\mathchar"218$}}
     \raise 2.0pt\hbox{$\mathchar"13E$}}}
\def\beq{\begin{equation}}
\def\eeq{\end{equation}}
\def\bce{\begin{center}}
\def\ece{\end{center}}
\def\bea{\begin{eqnarray}}
\def\eea{\end{eqnarray}}
\def\ben{\begin{enumerate}}
\def\een{\end{enumerate}}

\def\brr{\begin{array}}
\def\err{\end{array}}

\def\hmpc{\;h^{-1}{\rm Mpc}}

\def\nh1{n_{\rm HI}}

\def\p1dk{P_{\rm 1D}(k)}
\def\simlt{\mathrel{\spose{\lower 3pt\hbox{$\mathchar"218$}}
     \raise 2.0pt\hbox{$\mathchar"13C$}}}

\newbox\grsign \setbox\grsign=\hbox{$>$} \newdimen\grdimen \grdimen=\ht\grsign
\newbox\simlessbox \newbox\simgreatbox
\setbox\simgreatbox=\hbox{\raise.5ex\hbox{$>$}\llap
     {\lower.5ex\hbox{$\sim$}}}\ht1=\grdimen\dp1=0pt
\setbox\simlessbox=\hbox{\raise.5ex\hbox{$<$}\llap
     {\lower.5ex\hbox{$\sim$}}}\ht2=\grdimen\dp2=0pt

\title{Galaxy clustering in the Sloan Digital Sky Survey
(SDSS): A first comparison with the APM Galaxy Survey}

\author[Gazta\~naga]{
Enrique Gazta\~naga$^{1,2,3}$\\
$^1$ INAOE, Astrofisica, Tonantzintla, Apdo Postal 216 y 51, 
 Puebla 7200, Mexico \\
$^3$ Institut d'Estudis Espacials de Catalunya, ICE/CSIC,
Edf. Nexus-104-c/Gran Capita 2-4, 08034 Barcelona, Spain \\
$^2$ Departament de Fisica Fonamental, Universitat de
Barcelona, Diagonal 647, 08028 Barcelona, Spain}

\begin{document}

\maketitle

\begin{abstract}
  
  We compare the large scale galaxy clustering in the new SDSS early data
  release (EDR) with the clustering in the APM Galaxy Survey. We cut out pixel
  maps (identical in size and shape) from the SDSS and APM data to allow a
  direct comparison of the clustering.  Here we concentrate our analysis on an
  equatorial SDSS strip in the South Galactic Cap (EDR/SGC) with 166 $deg^2$,
  2.5 wide and 65 degrees long.  Only galaxies with Petrosian magnitudes $16.8
  <g' <19.8$ are included to match the surface density of the $17< b_J<20$ APM
  pixel maps (median depth of $\sim 400 \hmpc$).  Both the amplitude and shape
  of the angular 2-point function and variance turn out to be in very good
  agreement with the APM on scales smaller than 2 degrees (or $\la 15\hmpc$).
  The 3-point function and skewness are also in excellent agreement
  within a $90\%$ confidence level.  On one hand these results illustrate that
  the EDR data and SDSS software pipelines, work well and are suitable to do
  analysis of large scale clustering.  On the other hand it confirms that
  large scale clustering analysis is becoming "repeatable" and therefore that
  our conclusions for structure formation models seem to stand on solid
  scientific grounds.

\end{abstract}


\section{Introduction}

The Sloan Digital Sky Survey (SDSS) uses a dedicated 2.5 m telescope and a
large format CCD camera to obtain images of over 10,000 $deg^2$ 
 of high
Galactic latitude sky in five broad bands. The SDSS represents a new paradigm
of scientific project, where digital data is made available on-line to the
community via "virtual observatories". On June 2001 the SDSS collaboration
made an early data release (EDR) publicly available.  In this paper, we take
this opportunity to do a study of some large scale structure aspects of the
EDR.  We compare this new data with the APM Galaxy Survey (Maddox etal 1990),
which is based on 185 UK IIIA-J Schmidt photographic plates each corresponding
to $6 \times 6 $ square degrees on the sky limited to $b_J \simeq 20.5$ and
covering $b <-40$ and $\delta<-20$ degrees.  
These fields where scanned by the APM machine and carefully matched
using the plate overlaps.

The APM Survey has produced one of the best estimates of the angular galaxy
2-point correlation function $w_2(\theta)$ to date.  Its shape on large scales
led to the discovery of ``extra'' large-scale power, and gave early
indications for the paradigm shift out of the standard (Einstein-de Sitter)
CDM model (Maddox etal 1990).  Frieman \& Gazta\~naga (1999, FG99 hereafter)
estimated the 3-point galaxy correlation function in the APM Galaxy Survey and
its comparison with theoretical expectations (see also Peebles 1980, Fry 1984,
Juszkiewicz, Bouchet \& Colombi 1993, Fry \& Gazta\~naga 1993, Gazta\~naga
1994, Bernardeau 1994, Fosalba \& Gazta\~naga 1998, Buchalter, Jaffe \&
Kamionkowski 2000, Scoccimarro etal.  2001, and references therein).  For the
first time, the APM measurements extended to sufficiently large scales to
probe the weakly non-linear regime with a reliable Survey.  
The results are in good agreement with
gravitational growth for a model with initial Gaussian fluctuations.  They
also indicate that the APM galaxies are relatively unbiased tracers of the
mass on large scales and provide stringent constraints upon models with
non-Gaussian initial conditions.

Linear CCD devices provide more reliable
magnitudes than the traditional non-linear system of visual magnitudes.
Scanned plates should do better, but there has been extended discussions
regarding variable sensitivity inside individual Schmidt plates and
large-scale gradients in the APM survey calibration.  A CCD survey, such as
the SDSS, reduces these systematic effects
and could provide more accurate photometry and  luminosity function
estimators 
(eg Blanton etal 2001, see also Gazta\~naga \& Dalton 2000). Thus, a natural
question arrises: is the large scale clustering trace by the new SDSS digital
data compatible with the one measured in the old photographic APM scans?  As
there is no spatial overlap between these two data sets, we will compare the
statistical properties, such as correlation functions and moments of counts in
cells. The answer to this question will bring new evidence to our
understanding of structure formation in the lines mentioned above.

\section{SDSS sample and pixel maps}

We download data from the SDSS public archives using the SDSS Science Archive
Query Tool (sdssQT, http://archive.stsci.edu/sdss/software/).  We select
objects from an equatorial SGC (South Galactic Cap) strip 2.5 wide ($-1.25 <
DEC <1.25$ degrees.)  and 66 degrees long ($351 < RA < 56 $ degrees.)  which
seems to be homogeneous. This strip (SDSS numbers 82N/82S) corresponds to some
of the first runs of the early commissioning data (runs 94 and 125) with
variable seeing conditions (a few tenths of arc-second within scales of a few
degrees.
\footnote{See http://www-sdss.fnal.gov:8000/skent/seeingStatus.html}).
To avoid confusion with the EDR equatorial strip in the North (studied by the
SDSS collaboration, eg in Scranton et al 2001) we will refer to this strip as
EDR/SGC.

\begin{figure}
\centering{
{\epsfxsize=7cm \epsfbox{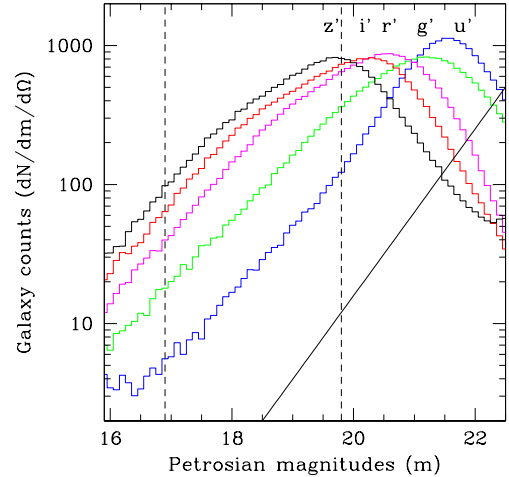}}}
\caption[Fig2]{\label{ncountsSGC} Galaxy density counts per 
magnitude and square deg. $dN/dm/d\Omega$ as a function of
Petrosian magnitude $z',i',r',g',u'$ (from left to right).}
\end{figure}

We first select all galaxies brighter than $u'=22.3, g'=23.3, r'=23.1,
i'=22.3, z'=20.8$, which corresponds to the SDSS limiting magnitudes for 5
sigma detection in point sources (York etal. 2000). Galaxies are found
from either the $1\times 1$, $2\times 2$ or $4\times 4$ binned CCD pixels 
and they are deblended by the SDSS pipeline (Lupton etal 2001).
There are about 375000
objects classified as galaxies in the EDR/SGC.  Figure \ref{ncountsSGC} shows
the number counts (surface density) 
for these 375000 galaxies as a function of the magnitude in
each band, measured by the SDSS modified Petrosian magnitudes $m_u'$, $m_g'$,
$m_r'$, $m_i'$ and $m_z'$ (see Yasuda et al 2001 for a discussion of the SDSS
counts).  Continuous diagonal lines show the $10^{0.6 m}$ expected for a low
redshift homogeneous distribution with no k-correction, no evolution and
no-extinction. The limiting magnitudes for detecting galaxies in this strip
seems to be closer to $m_u'<21.5$, $m_g'<21$, $m_r'<20.5$, $m_i'<20$ and
$m_z'<19.5$.

We next select galaxies with SDSS modified Petrosian magnitudes $16.8<
m_g'<19.8$. We choose $g'$ because this is the closest SDSS band to the APM
blue $b_J$.  The range $16.8< m_g'<19.8$ results from requiring that the SDSS
surface density in a 3 magnitude slice equals that of the APM in the $17 <
b_J< 20$ slice (ie $\simeq 300$ gal per $deg^2$). We have checked that small
departures from this range produce similar clustering results. There are about
54000 galaxies that match this criteria in the EDR/SGC.  Figure 
\ref{ncountsSGC}
shows that the number counts in this range of magnitudes (limit by the two
dashed vertical lines in the figure) follow well the $10^{0.6 m}$ relation
(continuous line), indicating that the sample selected is homogeneous and
completed to this magnitude. We do not apply any mask to the EDR/SGC
 sample or try to
make any correction for variable seeing or variable extinction across the
strip.
 Visually, there are no obvious holes or inhomogeneities in the pixel
maps at this depth.  We will assume that variations in extinction and seeing
do not affect clustering of the relatively bright galaxies we are 
considering (see a detailed study by Scranton etal 2001).

\begin{figure*}
\centering{
{\epsfxsize=14.cm \epsfbox{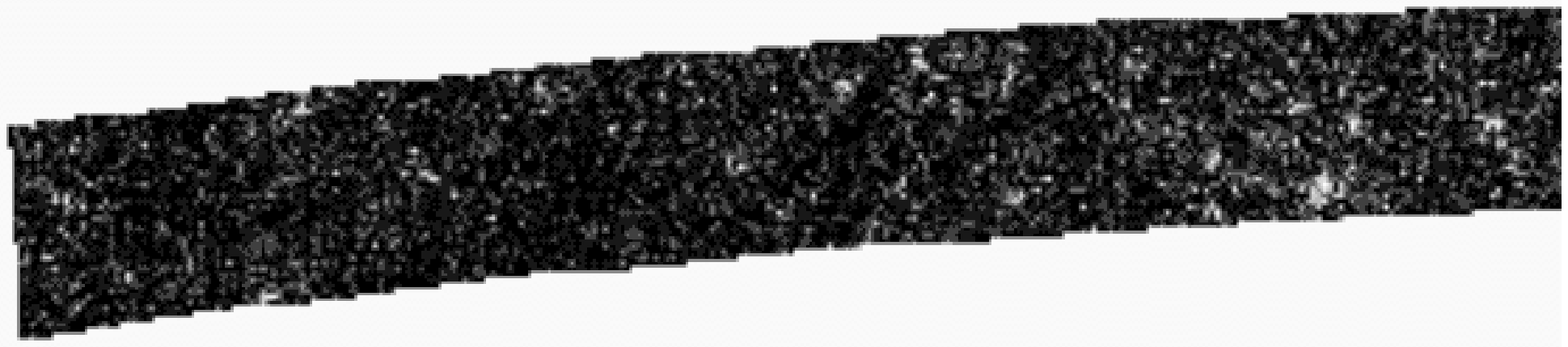}}
{\epsfxsize=14.cm \epsfbox{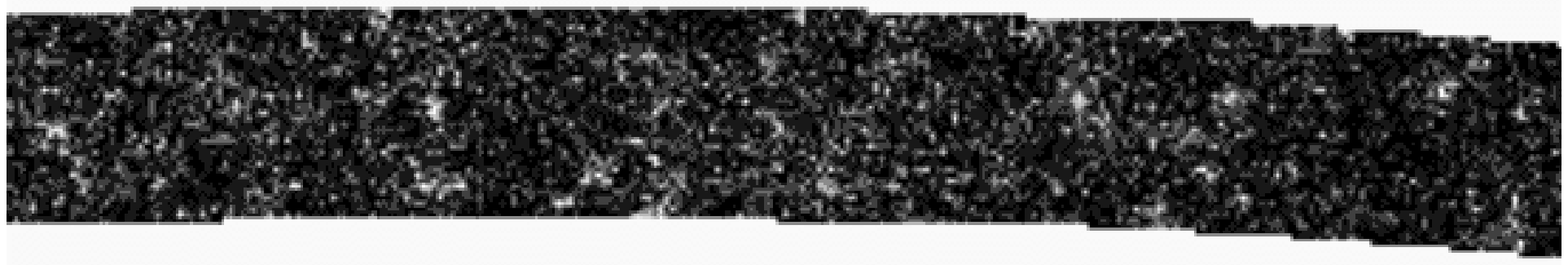}}
{\epsfxsize=14.cm \epsfbox{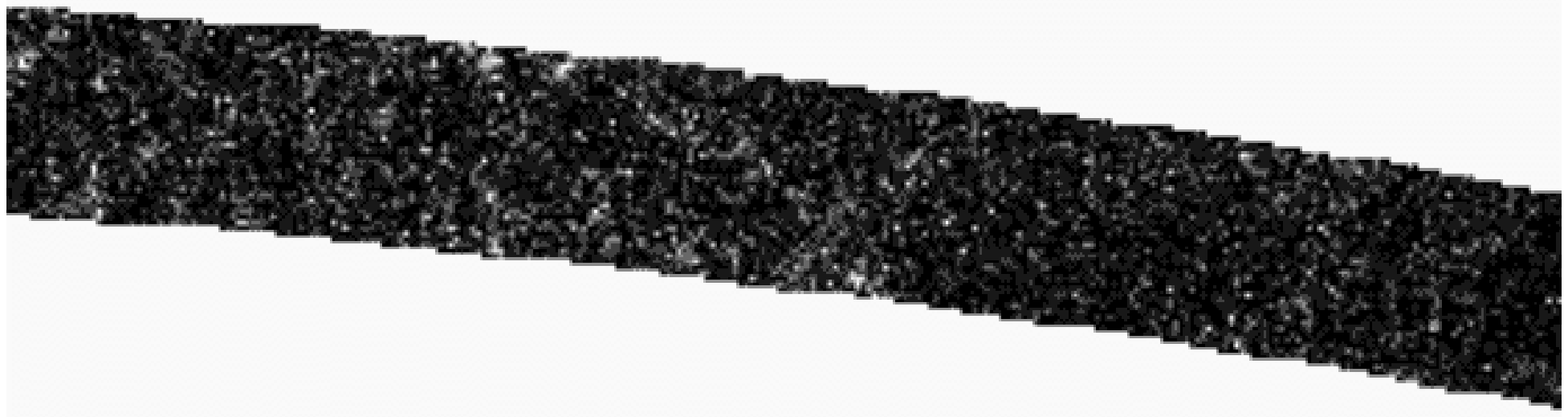}}
{\epsfxsize=14.cm \epsfbox{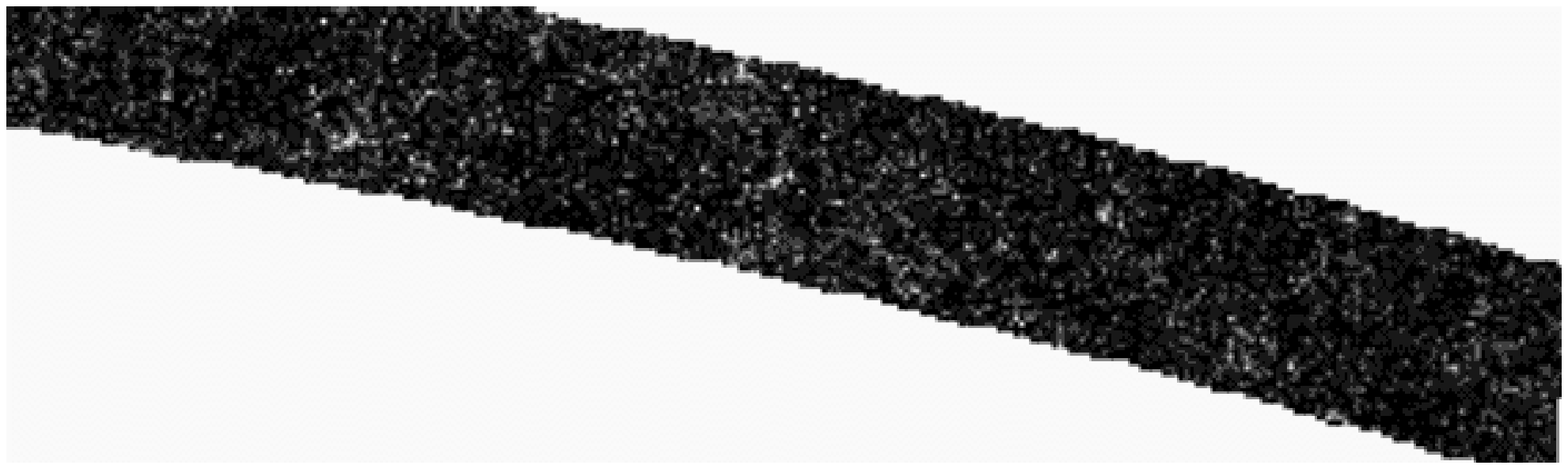}}
}
\caption[Fig1]{\label{Maps} Pixel maps showing number density of galaxies
increasing from 0 (black) to about 15 (white) galaxies per cells of $7$ 
arc-min.
The equatorial strip of $66$ degrees in RA and 2.5 degrees wide is cut in 4
overlapping pieces to fit the page.}
\end{figure*}

Finally, we produce equal area projection pixel maps of various resolutions
\footnote{Available on www.inaoep.mx/ $\tilde{ }$ gazta/Mapg18.9.cut.pgm.gz}, 
similar to those made
for the APM (see Plate 1 in Maddox etal 1990). 
 Figure \ref{Maps} shows
the  EDR/SGC pixel maps with a resolution of about $7$ arc-min. 
It is apparent 
that there are important surface density variations along the
strip, even on the largest scales.  One can also see filamentary structure and
a overall texture similar to that in the APM pixel maps (see Plate 1 in Maddox
etal. 1990) or the Lick maps (see cover of Peebles 1993).

\section{Clustering comparison}

Our goal is to compare the APM to the new SDSS data.  Out of the APM Survey we
considered a $17<b_J<20$ magnitude slice in an equal-area projection pixel map
with a resolution of $3.5$ arc-min, that covers over $4300$ $deg^2$ around the
SGC.  The APM can fit over 25 strips of similar area, shape and depth as the
SDSS strip shown in Figure 1. To study sampling and biasing effects on the
SDSS clustering estimators (due to the limited survey area), we have cut 10
strips out of the APM map. Each of these 10 sub-samples are identical in size
and shape to the EDR/SGC SDSS strip, and are separated from each other by few
degrees, so that they can be considered as independent.  This will allow us to
set $\simeq 90\%$ confidence intervals against the null hypothesis that the
EDR/SGC strip is compatible with the APM map. In all cases we correct the
clustering in the APM maps for a $5\%$ contamination of randomly merged
stars (see Maddox etal 1990), ie we scale fluctuations up by $5\%$
(see also Gazta\~naga 1994).

\subsection{Smoothed 1-point Moments}

\begin{figure}
\centering{
\epsfxsize=9cm \epsfbox{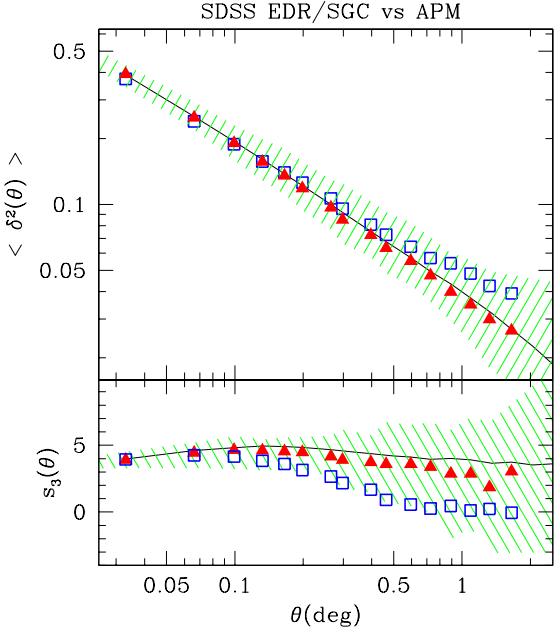}
}
\caption[Fig2]{\label{X2S3panel} The variance  $\bar{w_2}$ (top panel)
and reduced skewness (bottom panel)
as a function of angular smoothing $\theta$. Squares 
correspond to the SDSS EDR/SGC strip. 
The triangles and the shaded area correspond to 
the mean and $90\%$ confidence level in the values of
 10 APM sub-samples with same size and shape as the SDSS. The continuous line
corresponds to the whole APM map.}
\end{figure}

We first compare the lower order moments of counts in cells of variable size
$\theta$ (larger than the pixel map resolution).  We follow closely the
corresponding APM analysis in Gazta\~naga (1994, see also Szapudi etal 1995)
and use the same software and estimators here for the SDSS.  The top panel in
Fig. \ref{X2S3panel} shows the variance of fluctuations in density counts
$\delta \equiv \rho/{\bar{\rho}} -1$ smoothed over a scale $\theta$:
$\bar{w_2} \equiv < \delta^2(\theta) >$, which is plotted as a function of the
smoothing radius $\theta$.  The shaded region shows the $\simeq 90\%$
confidence interval for the SDSS results to be compatible to the APM map. This
shaded region brackets the minimum and maximum results at each scale in the 10
APM sub-samples that mimic the EDR/SGC. The individual results in each
subsample are strongly correlated so that the whole curve for each subsample
scales up an down inside the shaded region, ie there is a strong covariance at
all separations due to large scale density fluctuations.  Triangles correspond
to the mean of the 10 APM sub-samples, which are slightly bias down as
compared to the whole APM (continuous line). The SDSS results (open squares)
are all within the $90\%$ confidence region set by the APM.  Because of the
strong covariance mentioned above, we do not expect the SDSS values to scatter
around the APM values, but rather we expect the whole curved to be shifted, as
shown in the Figure. This is also the case for the other statistical
measures presented below.

The extend of the shaded area would narrow if we increase 
the total solid angle of the subsamples.
Numerical simulations show that the corresponding error-bar is 
about a factor of 4-6 smaller  for a survey as large as the APM.

The bottom panel in 
Fig. \ref{X2S3panel} shows the corresponding comparison for the normalized
angular skewness:
\beq
s_3(\theta) \equiv {<\delta^3(\theta)>\over{<\delta^2(\theta)>^2}} \equiv {\bar{w_2}(\theta)
\over{\bar{w_3}(\theta)}}
\eeq
Again the mean of the individual APM zones (triangles) 
does not equal the whole APM estimation (continuous line).
This is due to estimation (or ratio) bias (see Hui \& Gazta\~naga 1999).
The SDSS EDR/SGC 
data on $s_3$ shows an excellent agreement with the APM at the
smaller scales (in contrast with the EDSGC results, see Szapudi \& Gazta\~naga
1998).  On larger scales the SDSS values are smaller, but the discrepancy is
not significant within the $90\%$ region and given the strong covariance
of individual subsamples.
 Similar results are found for higher
order moments.  As we approach the scale of $2$ degrees, the width of our
strip, it becomes impossible to do counts for larger cells.
It is therefore interesting to study the correlation
functions, which should be less affected by boundary effects (although
sampling effects will still be important).

\subsection{N-point Correlation functions}

We next study the angular 2-point, $w_2\equiv <\delta_1 \delta_2>$, and
3-point, $w_3 \equiv <\delta_1 \delta_2 \delta_3 >$, correlation functions.  
Here we follow closely the notation, estimators and software used in FG99.
The 3-point function is normalized as:
\beq
q_3 \equiv {
w_3(\theta_{12}, \theta_{13}, \theta_{23}) \over
w_2(\theta_{12})w_2(\theta_{13})+w_2(\theta_{12})w_2(\theta_{23})+
w_2(\theta_{13})w_2(\theta_{23})}
\label{Qp}
\eeq
where $\theta_{12}$, $\theta_{13}$ and $\theta_{23}$ correspond to
the sides of the triangle form by the 3 angular positions
of $\delta_1 \delta_2 \delta_3$. Here we will
consider isosceles triangles, ie $\theta_{12}=\theta_{13}$,
so that $q_3=q_3(\alpha)$ is given as a function of the interior 
angle $\alpha$ which
determines the other side of the triangle $ \theta_{23}$ (FG99).
We also consider
the particular case of the collapsed configuration  $\theta_{23}=0$,
which corresponds to $ <\delta_1 \delta_2^2>$ and is normalized in slightly
different way (see also Szapudi \& Szalay 1999):
\beq
c_{12} \equiv {<\delta_1 \delta_2^2>\over{<\delta_1 \delta_2><\delta_1^2>}}
\simeq 2 q_3(\alpha=0) ~.
\eeq

\begin{figure} 
\centering{
{\epsfxsize=9cm \epsfbox{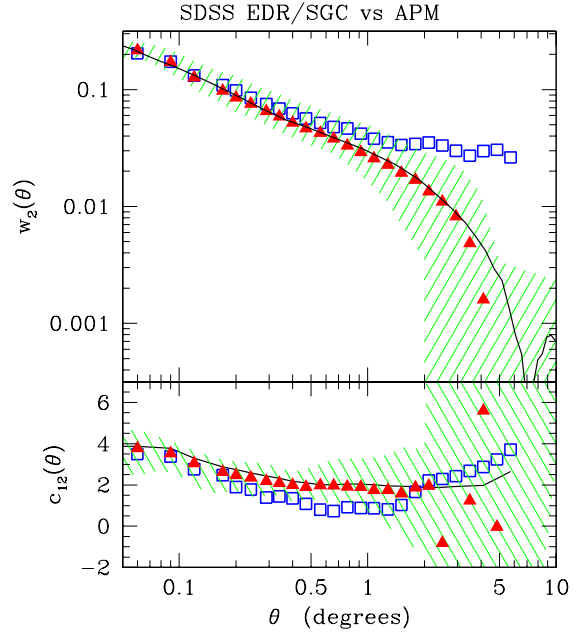}}
}
\caption[Fig2]{\label{w2c12panel}  Same as Fig.\ref{X2S3panel} for
the 2-point function (top panel)  and the collapsed 3-point function, $c_{12}$,
as a function of scale $\theta$.}
\end{figure}

Top panel in Fig.\ref{w2c12panel} shows the 2-point correlation as a function
of the separation $\theta$ between points (here small pixel cells).  Results
from both surveys agree remarkably well up to 1 degree. In a similar way as to
what happened with the variance (ie Fig.\ref{X2S3panel}) the SDSS EDR/SGC
sample shows a flatting of the $w_2$ slope at large scales. At scales bigger
than $\simeq 3$ degrees this discrepancy becomes significant and it is
incompatible with the APM subsamples with a confidence level higher that
$90\%$.  This deviation could be due to systematic gradients in the SDSS raw
data that we are using (eg due to large seeing variation across the strip, see
\S 2 above). But it could also be due to an underestimation of the sampling
errors in the APM (eg due to smoothing of the APM calibration on the scale of
the subsamples).  
We need to consider larger and better calibrated areas to
confirm or refute with the SDSS the features in APM galaxy power spectrum
shape discussed in Gazta\~naga \& Baugh (1998, and references therein) and
Gazta\~naga \& Juszkiewicz (2001).  At smaller scales, the agreement is
excellent and the measured shape of the SDSS 2-point function confirms the
idea that galaxies are 'anti-bias' with respect to the $\Lambda$CDM model (eg
Gazta\~naga 1995, Jenkins etal 1998, and references therein).

\begin{figure} 
\centering{
{\epsfxsize=7cm \epsfbox{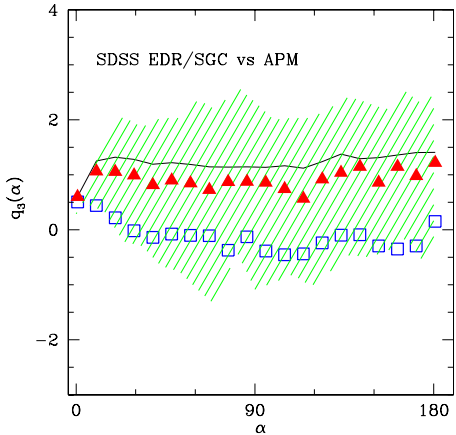}}
}
\caption[Fig2]{\label{q3sdss}  Same as Fig.\ref{X2S3panel} for
3-point function $q_3(\alpha)$.}
\end{figure}

Bottom panel of Fig.\ref{w2c12panel} and Fig.\ref{q3sdss} shows the reduced
3-point function $q_3$ for different configurations.  Fig.\ref{q3sdss}
compares $q_3(\alpha)$ for isosceles triangles of side
$\theta_{12}=\theta_{13}=0.5$ degrees.  Bottom panel of Fig.\ref{w2c12panel}
shows the collapsed case $c12 \simeq 2 q_3(\alpha=0)$ as a function of the
other triangle side $theta=\theta_{12}=\theta_{13}$.  The reduced 3-point
function in Fig.\ref{q3sdss} is close to zero and the signal seems stronger on
smaller scales.  In all cases, the SDSS results are within the $90\%$
confidence of the APM values.  Note again the bias between the mean of the 10
APM strips (triangles) and the whole APM (continuous line), and also the
strong covariance across the curves.  As mentioned above both of these results
are expected and due to (very) large scale density fluctuations.

\section{Discussion and conclusions}

We have presented a rigorous $90\%$ confidence test for the hypothesis that
the clustering in the SDSS EDR/SGC strip is compatible with the one in the APM
data.  The test is passed for clustering at scales smaller than 2 degrees
(which corresponds to $\simeq 15 Mpc/h$), and seems to fail (at $90\%$
confidence) at larger scales. This could be due to large scale variations in
the seeing (on scales of a few degrees) present in the early SDSS
commissioning data (EDR/SGC) that we are using (see \S2), but they could also
reflect the fact that the APM sampling errors are not accurate at this level. In
principle, it is possible to correct for these seeing gradients by using the
appropriate mask (see Scranton etal 2001), but we have not attempted to do this
here.

The agreement between these two catalogues, on scales smaller than 2 degrees,
is not trivial: correlation functions are quite sensitive to systematic
effects and clustering for galaxies of fainter magnitudes or different colors
is quite different. Moreover, we have not rescale or match the amplitudes of
the two catalogues: the agreement comes naturally when we just select galaxies
with the same colors, magnitude range and similar surface density 
in both samples.  It is hard to belive that this is just a coincidence.  
These results rather illustrates that EDR data and software pipelines 
from SDSS work well and are suited to do analysis of large scale clustering.

We find it remarkable that two totally independent surveys, separated in space
(ie angular position and spectral filtering) and time (over 10 years in
technological and software development) provide very similar results for the
large scale angular clusterings trace by galaxies.  This is true both visually
and by accurate statistical comparisons.  It illustrates that large scale
clustering analysis is becoming "repeatable" and therefore that our
conclusions could stand on solid scientific grounds.  In particular, one
conclusion that we have been able to confirm in this work, is that the higher
order correlations in galaxy samples indicate that gravitational growth from
Gaussian initial conditions is responsible for the hierarchical structures we
see in the sky (eg FG99, Scoccimarro etal 2001, and references therein).  When
more of the SDSS data is analyzed we can hope to make much higher precession
testing of our models for structure formation (eg Tegmark etal 1998; Colombi
etal 2000).

\section*{Acknowledgments}

I acknowledge support by grants from IEEC/CSIC and DGI/MCT BFM2000-0810.
Funding for the creation and distribution of the SDSS Archive has been
provided by the Alfred P. Sloan Foundation, the Participating Institutions,
the National Aeronautics and Space Administration, the National Science
Foundation, the U.S. Department of Energy, the Japanese Monbukagakusho, and
the Max Planck Society. The SDSS Web site is http://www.sdss.org/.



\end{document}